\documentclass[12pt]{spieman}  
\usepackage{amsmath,amsfonts,amssymb}
\usepackage{graphicx}
\usepackage{setspace}
\usepackage{tocloft}

\usepackage{xurl}

\title{\textit{In vivo} fundus imaging and computational refocusing with a diffuser-based fundus camera}

\author[a]{Corey Simmerer}
\author[a]{Marisa Morakis}
\author[b]{Lei Tian} 
\author[a,c,d]{Lia Gomez Perez}
\author[e]{T.Y. Alvin Liu}
\author[a,e,*]{Nicholas J. Durr}
\affil[a]{Johns Hopkins University, Department of Biomedical Engineering, Baltimore, Maryland, USA, 21218}
\affil[b]{Boston University, Department of Electrical and Computer Engineering, Boston, Massachusetts, USA, 02215}
\affil[c]{Massachusetts General Hospital, Wellman Center for Photomedicine, Boston, Massachusetts, USA, 02114}
\affil[d]{Harvard-MIT Health Sciences and Technology, Cambridge, Massachusetts, USA, 02142}
\affil[e]{Wilmer Eye Institute, Department of Ophthalmology, Baltimore, Maryland, USA, 21218}

\cftpagenumbersoff{figure}
\cftpagenumbersoff{table} 
\begin{document} 
\maketitle

\begin{abstract} \\
\textbf{Significance:}  Access to diagnostic eye care could be expanded with high-throughput and easy-to-use tools. Phase mask-based imaging may improve the fundus camera by enabling computational refocusing with no moving parts. While phase mask-based imaging has been demonstrated in a model eye, this approach has not been shown \textit{in vivo}. \\
\textbf{Aim:} A computational fundus camera was designed, constructed, and evaluated with the goal of determining the feasibility and performance of phase mask-based computational imaging of the \textit{in vivo} fundus.\\
\textbf{Approach:} A holographic diffuser was introduced in a modified commercial fundus camera at a plane conjugate to the ocular pupil, resulting in a linear and shift-invariant point spread function that varies with refractive error. The image could be digitally refocused across a range of $\geq\pm$ 10 diopters of defocus error. The device was tested for ocular safety, and a human imaging pilot study was performed. \\ 
\textbf{Results:} The device captured and digitally refocused color human fundus images. The field of view was $\geq$35 degrees and resolution was 7.7-9.6 line pairs per mm. \\
\textbf{Conclusions:} We present the first \textit{in vivo} diffuser-based fundus images, demonstrating the feasibility of computational imaging for ocular diagnostics. 
\end{abstract}


{\noindent \footnotesize\textbf{*}Nicholas J. Durr,  \linkable{ndurr@jhu.edu} }

\begin{spacing}{2}   

\section{Introduction}
\label{sect:intro} 

The comprehensive eye examination is the foundational procedure to assess and address eye disease. A major component of this is the imaging of the fundus, which is the posterior surface of the eye where the retina and optic nerve head is located \cite{hoang_adult_2019}. Fundus examinations screen for diseases such as diabetic retinopathy, glaucoma, and age-related macular degeneration, which are leading causes of vision impairment and blindness globally \cite{flaxman_global_2017}. There is an important global need to improve the accessibility of eye examinations to prevent and treat vision loss and blindness \cite{durr_unseen_2014}. Even in high-risk populations in high-resource countries, fundus examinations are underperformed, resulting in unmanaged vision loss \cite{bragge_screening_2011,chin_nonmydriatic_2014}. Currently, the digital fundus camera is the primary tool used to perform fundus examinations \cite{bernardes_digital_2011}. However, they are expensive and difficult to focus and align.  Recently developed portable fundus cameras aim to alleviate some barriers to access \cite{yao_developing_2022}. However, these handheld devices using traditional optics are still challenging to align and focus. Significant barriers to access such as the high cost of equipment, the high skill level required to perform these sensitive measurements, and the shortage of trained clinicians and technicians in remote areas, restrict the utilization of fundus examination \cite{panwar_fundus_2016, bascaran_effectiveness_2021,dunn_optimising_2023}. Creating a device that simplifies fundus image capture has great potential to reduce costs, increase throughput, and deskill the comprehensive eye examination.

The emerging field of computational lensless imaging promises new capabilities for imaging with a standard image sensor, such as single-shot 3D imaging \cite{antipa_diffusercam_2018, boominathan_phlatcam_2020}, computational refocusing\cite{antipa_single-shot_2016}, and lightfield imaging \cite{cai_lensless_2020}. One method of computational imaging uses an encoding mask instead of a lens to multiplex plenoptic information onto the image sensor\cite{boominathan_recent_2022}. Prior knowledge of the optical properties of the mask then enables computational reconstruction of the captured lightfield. One such encoding mask is a holographic diffuser, as demonstrated by Antipa et al. \cite{antipa_single-shot_2016}. The diffuser produces a sharp caustic pattern as a point spread function (PSF) that is nearly shift-invariant for lateral movements of a point, while axial movements cause scaling. Prior characterization of the PSF enables image reconstruction from diffuser measurements. The axial dependence of the PSF enables 3D imaging and computational refocusing. These advancements are promising, but applications remain relatively unexplored. Computational imaging may be advantageous in a fundus camera, which would allow the focusing process to take place after image capture.

Li et al. recently proposed the diffuser-based computational imaging funduscope \cite{li_diffuser-based_2020}. They used a 4-$f$, infinite-conjugate system, relaying the wavefront emerging from a model eye pupil to a diffuser placed in front of the image sensor. They demonstrated grayscale imaging and computational refocusing capabilities with varying levels of defocus error on a model eye. An additional motivator for their diffuser-based computational imaging funduscope is the potential to combine fundus imaging and aberrometry into one device. Using a very similar optical configuration, McKay et al. \cite{mckay_large_2019} created a diffuser-based wavefront sensor that could be used to perform aberrometry over a very large dynamic range. The combination of aberrometry (for eyeglass prescriptions) and fundus imaging (for disease screening) in one easy-to-use device is an enticing opportunity that could be realized with computational ophthalmic imaging. This paper details our progress in color computational fundus imaging \textit{in vivo}. We incorporated modern fundus camera design principles and expanded upon the methods used by Li et al. \cite{li_diffuser-based_2020} to address unique challenges to imaging \textit{in vivo} and implement color imaging. We conducted a detailed evaluation of the imaging resolution and field of view of the system. Finally, we test the device for human eye light safety and present the first diffuser-based \textit{in vivo} fundus images, including computational refocusing. This work represents a significant step forward for this novel application of computational imaging, demonstrating the feasibility of \textit{in vivo} fundus imaging with a phase mask.


\section{Methods}
\subsection{Overview}

\begin{figure}
    \begin{center}
    \begin{tabular}{c}
    \includegraphics[height=6cm]{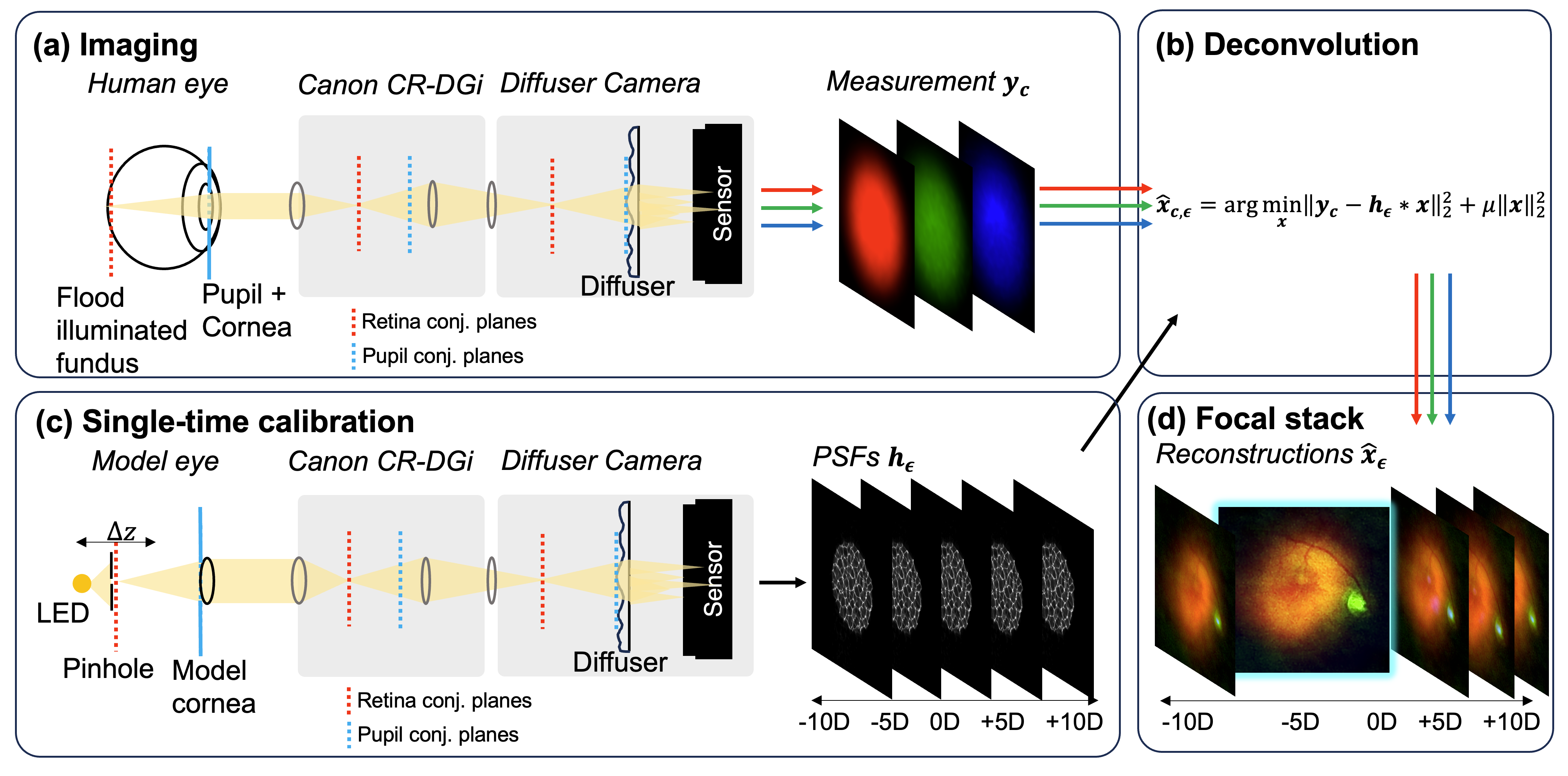}
    \end{tabular}
    \end{center}
    \caption{ \label{fig:overview}
    Overview of the computational fundus camera. (a) Measurements of the fundus are captured by connecting a relay lens, diffuser, and image sensor to a commercial fundus camera (Canon Cr-DGi). Three consecutive measurements, $\boldsymbol{y_c}$, using red, green, and blue light illumination are taken with the diffuser in a conjugate plane to the pupil. Dashed lines indicate conjugate planes. (b) The three color channels are independently deconvolved with each PSF in the calibration stack. (c) The system is calibrated by directly measuring the PSF over a range of axial displacements ($\Delta z$) from the focal plane of the model cornea, inducing a range of defocus errors, $\epsilon$. (d) An image is reconstructed from each color channel for each PSF and the sharpest image is selected from this stack.} 
\end{figure}

A major challenge for \textit{in vivo} imaging arises from the requirement that the illumination and imaging optical pathways pass coaxially through the pupil. The cornea is much more reflective than the fundus, and the strong corneal reflections from the illumination light must be blocked from passing back through to the image sensor \cite{dehoog_optimal_2008}. In the computational imaging funduscope by Li et al. \cite{li_diffuser-based_2020}, this is achieved using crossed polarizers. This eliminates much of the backreflections, but also limits the signal from the fundus. For \textit{in vivo} imaging, the illumination power must be limited to within safe limits, which requires maximizing the amount of light collected from the fundus, especially for computational imaging where signal-to-noise ratio (SNR) is important. Therefore, to demonstrate feasibility of computational fundus imaging \textit{in vivo}, we used a modified commercial non-mydriatic fundus camera as the foundation for our system. Conventional clinical fundus cameras use an alternative optical design in a non 4-$f$ configuration with a holed mirror to eliminate the need for polarizers while keeping illumination and imaging pathways coaxial. This ensures a high SNR and short exposure time for \textit{in vivo} imaging. Our system utilizes a diffuser-based camera inspired by Antipa et al. \cite{antipa_diffusercam_2018}. An overview of the computational imaging workflow and system layout is shown in Fig. \ref{fig:overview}. Similar to the layout presented by Li et al. \cite{li_diffuser-based_2020}, a holographic diffuser is placed conjugate to the ocular pupil. Then, an image sensor is positioned at an axial distance behind the diffuser that optimizes the sharpness of the PSF. The PSF is approximately linear and shift-invariant (LSI) in the transverse axis, meaning that the sensor measurement can be represented by a convolution of the object at a specific axial plane with the PSF corresponding to that plane. The PSF depends on the axial distance of the fundus behind the ocular lens and scales with this defocus error \cite{antipa_diffusercam_2018}. These PSFs are measured a single time to calibrate the system. For color fundus imaging, three consecutive images are collected in red, green, and blue illumination on a monochromatic sensor. The resulting image is deconvolved with the PSF calibration stack using a regularized inverse filter, and the most focused image is selected from the resulting focal stack.

\subsection{Optical hardware}
\begin{figure}
    \begin{center}
    \begin{tabular}{c}
    \includegraphics[height=8cm]{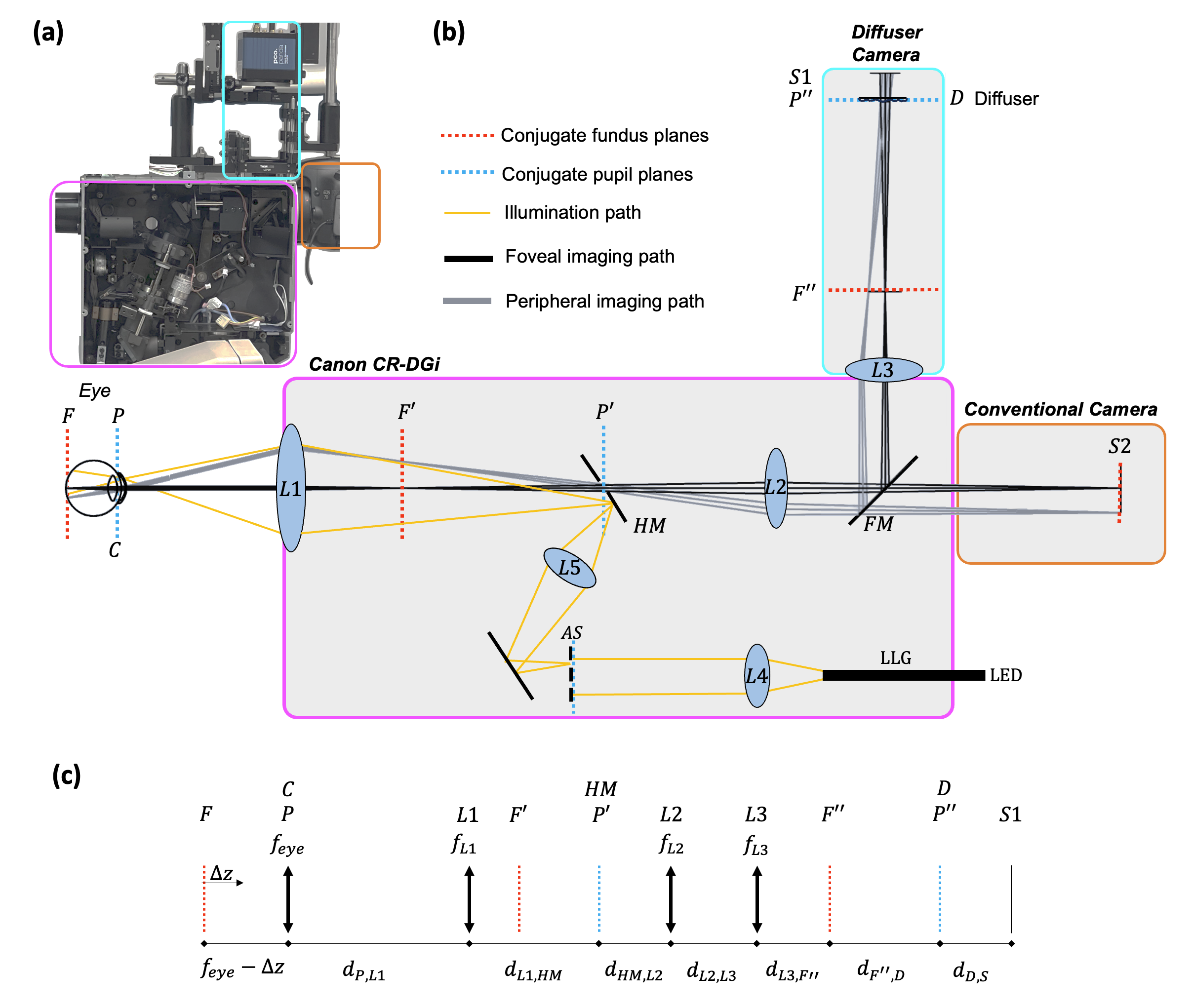}
    \end{tabular}
    \end{center}
    \caption{ \label{fig:optics}
    Optical layout of the computational fundus camera. (a) Photo of the system with the main subsystems highlighted. (b) Annotated Zemax model of the system using paraxial lenses. Flood illumination of the fundus is provided with LEDs coupled through a liquid light guide (LLG). The annular slit (AS) is imaged to the outer edges of the pupil of the eye. Remitted light from the fundus is collected from the center of the pupil that is conjugate with a holed mirror (HM). A conventional image of the fundus is formed on S2. For computational imaging, the flip mirror (FM) directs light through L3, which forms a conjugate pupil plane at the diffuser. (c) Reduced model of the diffuser imaging pathway with conjugate planes and distance variables defined. Values for these parameters are in Tables \ref{tab:parts} and \ref{tab:vars}. Defocus error may be applied to the reduced eye by adjusting the distance between $F$ and $P$ by $\Delta z$.} 
\end{figure}

We used an existing commercial fundus camera (CR-DGi, Canon) as the platform for our novel computational fundus camera. The illumination and imaging pathways of the commercial system are optimized for non-mydriatic use and corneal backreflections are blocked without using polarizers, increasing throughput. The Canon CR-DGi also has a flip mirror to switch between the infrared alignment camera and the imaging camera. We removed the infrared alignment camera and replaced it with the diffuser-based computational camera ("Diffuser Camera" in Fig. \ref{fig:optics}) while keeping the conventional imaging camera (EOS 7D, Canon) in place. This enables switching between conventional and diffuser imaging.

The illumination pathway of the system is nearly identical to the original commercial fundus camera with the internal lenses in the housing unchanged (L1, L4, L5 in Fig. \ref{fig:optics}). This optical design simultaneously illuminates and images the fundus through the pupil. The illumination and imaging paths must be coaxial and both pass through the cornea, which is much more reflective than the fundus \cite{berendschot_fundus_2003, dehoog_fundus_2009}. The Canon CR-DGi has well-designed optics for this task, so we did not alter this pathway. The original flash lamp was replaced with a liquid light guide (LLG03-4H, Thorlabs) connected to an LED light source (Lumen 1600-LED, Prior Scientific) with multiple selectable wavelengths in the visible and NIR range. The illumination is collimated with L4 (Fig. \ref{fig:optics}) and is passed through an annular slit (AS). The image of the ring is imaged onto the holed mirror (HM) with L5 and then imaged onto the pupil with the objective lens L1. The illumination ring is centered on the periphery of the pupil and results in flood illumination of the fundus while leaving the inner aperture available for back-reflection-free imaging. 

The light remitted from the fundus propagates back through the pupil and cornea, is relayed through the aperture in HM with L1 and L2, and is either imaged to a conventional camera or directed to the diffuser system via a flip mirror (FM). For conventional fundus imaging, L2 is adjusted with axial movements to accommodate for defocus error (myopia or hyperopia), whereas for diffuser imaging, L2 is fixed. For the diffuser imaging path, the flip mirror reflects the light through L3 (AC254-100-A, $f$=100 mm, Thorlabs) to create a conjugate pupil plane at a 0.5$^\circ $ holographic diffuser (\#47-988, Edmund Optics). An image sensor (pco.panda 4.2 bi UV, Excelitas Technologies) is placed 11 mm behind the diffuser, which was found to be the location that produces the sharpest caustic PSF. We modeled this path in Zemax using paraxial lenses to produce the ray diagram in Fig. \ref{fig:optics}. A full list of optical parameters of the components is in Table \ref{tab:parts} in the appendix.

\subsection{Reconstruction}
To computationally reconstruct the image from the diffuser measurement, we model the object (the fundus) as a surface in 3D space at some axial depth behind the cornea associated with some defocus error, $\epsilon$. We assume that the fundus is approximately flat across the roughly 45 field of view of our system, which is reasonable because the fundus has relatively little curvature at this scale. Consistent with other reports and our PSF measurements, our system is approximately LSI in each transverse plane \cite{antipa_diffusercam_2018,li_diffuser-based_2020}. The PSF for a given plane with an associated defocus error is $\boldsymbol{h_\epsilon}$. Because the system is LSI, we model the measurement, $\boldsymbol{y}$, of a fundus, $\boldsymbol{x}$, with convolution.

We also assume that the illumination field is flat and that the PSF does not vary significantly with wavelength. The object at a given color of illumination is represented by $\boldsymbol{x_c}$, and the resulting measurement by $\boldsymbol{y_c}$. The noise model is additive white Gaussian noise, $\boldsymbol{n}$.

\begin{equation}
\boldsymbol{y_c}=\boldsymbol{h_\epsilon}*\boldsymbol{x_c}+ \boldsymbol{n}
\end{equation}

The reconstruction task is to estimate the underlying fundus,  $\boldsymbol{x_c}$, given the measurement $\boldsymbol{y_c}$. We estimate the fundus image for each color at each defocus error, $\boldsymbol{\hat{x}_{\epsilon, c}}$, using Tikhonov regularization.

\begin{equation}
\boldsymbol{{\hat{x}_{\epsilon, c}}} = \arg \min_{\boldsymbol{x}} \lVert \boldsymbol{y_c} - \boldsymbol{h_\epsilon}*\boldsymbol{x} \rVert_2^2 + \mu \lVert \boldsymbol{x} \rVert _2^2
\label{eq: minproblem}
\end{equation}

Regularization strength is controlled by $\mu$, which is adjustable. The first term minimizes $L_2$-norm of the difference between the model and the measurement. The second term regularizes the output by penalizing the $L_2$-norm of the estimate. This convex optimization problem can be solved efficiently in the Fourier domain \cite{murli_wiener_1999}. $\boldsymbol{\hat{X}_{\epsilon, c}}$ is the Fourier transform of the estimated signal, $\boldsymbol{{\hat{x}_{\epsilon,c}}}$. $\boldsymbol{H_\epsilon}$ and  $\boldsymbol{Y_c}$ are the Fourier transforms of $\boldsymbol{h_\epsilon}$ and  $\boldsymbol{y_c}$, respectively.

\begin{equation}
\boldsymbol{{\hat{x}_{\epsilon,c}}} = \mathcal{F}^{-1} \{ {\boldsymbol{\hat{X}_{\epsilon, c}}} \} =
\mathcal{F}^{-1} \left\{  \frac{\boldsymbol{H_\epsilon^*}}{\lvert \boldsymbol{H_\epsilon} \rvert^2+\mu}   \boldsymbol{Y_c}\right\}
\label{eq: inversefilter}
\end{equation}

We implement this reconstruction using the fast Fourier transform (FFT) and find that reconstructions take under one second to compute for 16-bit 4.2 megapixel images on a Dell Precision 5820 Workstation with a 14-core Intel Xeon W-2275 3.3 GHz CPU.

The eye's defocus error, and thus the corresponding $\boldsymbol{h_\epsilon}$, is unknown at the time of image capture. After image capture, a focal stack of reconstructed fundus images is created by deconvolving the measurements, $\boldsymbol{y_c}$, with each PSF in the set of recorded $\boldsymbol{h_\epsilon}$. This is done for each color channel, producing a focal stack of color reconstructions. The focal stack can then be searched for the sharpest image manually or with an autofocusing algorithm. 

The regularization strength, $\mu$, was chosen empirically. A stack of reconstructions using the correct PSF and a range of $\mu$ from 0.001 to 0.025 was created for each color channel. Then, the image that subjectively balances sharpness with noise in each color channel was chosen. For red, this was chosen to be $\mu=0.007$, green used $\mu=0.002$, and blue used $\mu=0.02$. The use of different $\mu$ for the different channels can be justified since the color channels have different signal-to-noise ratios. The high contrast green channel requires less regularization than the low-signal, low-contrast blue channel.

\subsection{Calibration}

\begin{figure}
    \begin{center}
    \begin{tabular}{c}
    \includegraphics[height=8cm]{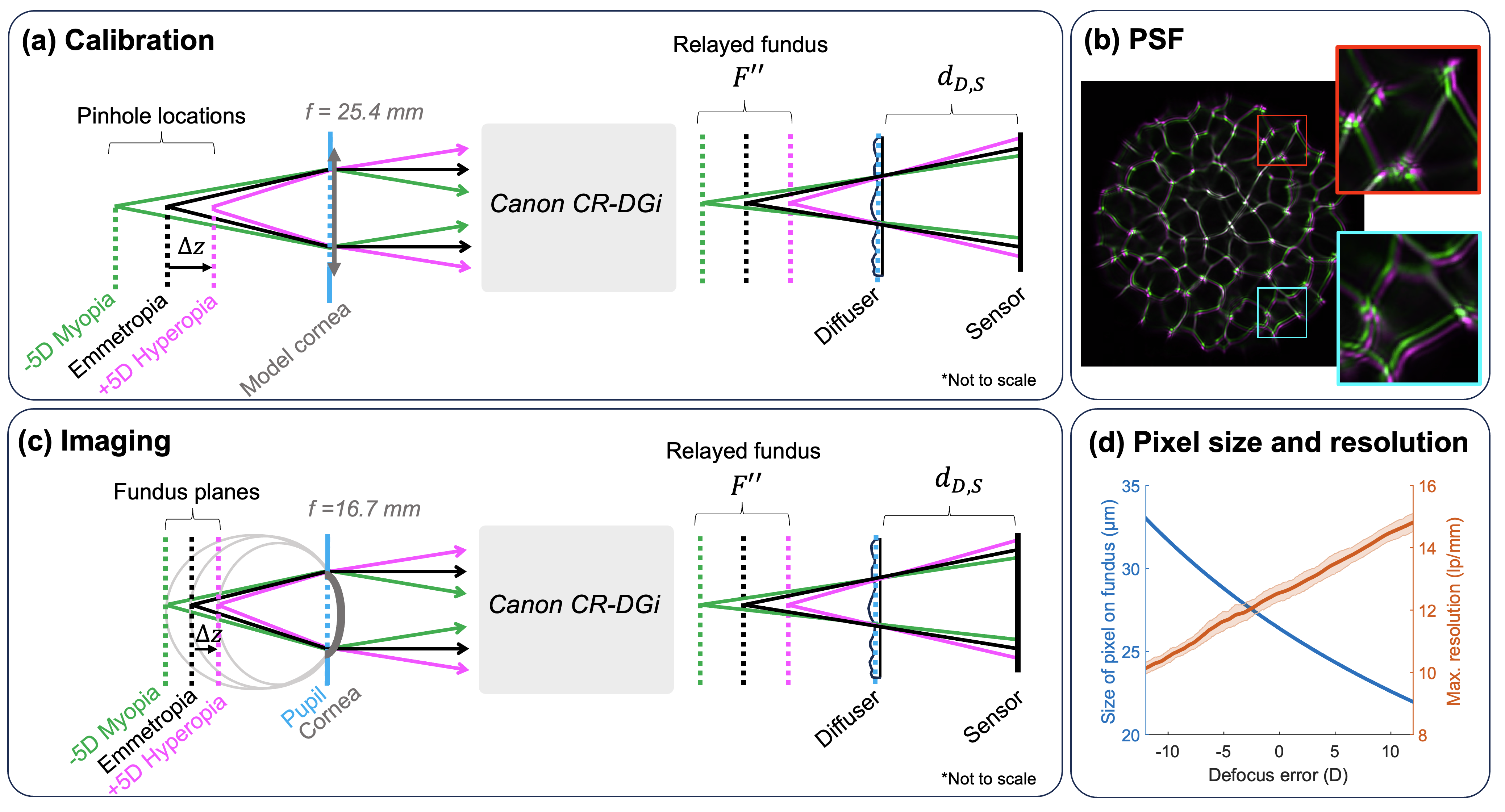}
    \end{tabular}
    \end{center}
    \caption{ \label{fig:vergence}
    Calibration overview. (a) The calibration of the system involves measuring PSFs of the system for a range of defocus errors, $\epsilon$. This is done by axially translating a 20-micron pinhole over a range of $\Delta z$ from the backfocal plane of the model cornea. (b) Measured PSFs corresponding to -5D and +5D of defocus error (green and magenta, respectively), demonstrating the magnification difference. (c) An eye with a defocus error is modeled as having a fundus that is offset from the backfocal plane of the cornea by some $\Delta z$. The wavefront remitted from the pupil has a vergence dependent on the defocus error and is relayed to the diffuser. This vergence scales the PSF (d) The reconstruction pixel size and the theoretical resolution as a function of defocus error. The theoretical resolution is determined by the maximum frequency component of the PSF with a magnitude greater than the noise floor of the camera.} 
\end{figure}

Calibration of the system involves measuring a set of PSFs spanning a range of defocus errors. This is a single-time calibration process which takes approximately 15 minutes to measure with a motorized translation stage. The calibration can then be used to deconvolve all subsequent images taken with the system. The stage (LTS300, Thorlabs) is attached to a 20-micron diameter pinhole in front of a bright white LED (SLS-0300-C, Mightex). To simulate a point source on the retina of an emmetropic (no refractive error) eye, the pinhole is placed a focal length behind a convex lens acting as a model cornea (LB1761-A, f=25.4 mm, Thorlabs), as diagrammed in Fig. \ref{fig:vergence}. The recorded signal on the image sensor is the PSF,  $\boldsymbol{h_\epsilon}$, for the defocus error associated with the sampled $\Delta z$. We sampled PSFs from $\Delta z = -12 $mm to $\Delta z = +6$ mm with a sampling period of 50 microns, corresponding to defocus errors from -12 to +12 diopters (D). 

The PSF is dependent on defocus error, $\epsilon$, which can be computed from the focal length of the model eye lens, $f_{\mathrm{model}}$, and the distance of the object from the focus of the lens, $\Delta z$. The defocus error (in diopters) is the difference in optical powers:
\begin{equation}
    \epsilon = \frac{1}{f_{\mathrm{model}}-\Delta z}-\frac{1}{f_{\mathrm{model}}}
    \label{eq:refractive error}
\end{equation}

\subsection{Theoretical resolution and FOV}
The magnification of the system can be computed with ray matrix analysis.  We use a reduced eye model in this analysis, assuming that the eye is a thin lens with a power of 60 diopters, $f_{\mathrm{eye}} = 100/6 \: \mathrm{mm}$. The distance between the cornea and fundus of an eye with a defocus error caused by an axial displacement of the fundus is $f_{\mathrm{eye}}-\Delta z$. The system from the cornea to L3 is fixed and can be represented by a ray matrix, $\bigl[ \begin{smallmatrix}
    A & B \\ C & D
\end{smallmatrix}\bigr] $. The system matrix from the fundus, $F$, to the final conjugate fundus plane, $F''$, is represented by the matrix $\mathbf{H}$.  The distance from L3 to the final conjugate fundus plane, $F''$, is $d_{L3,F''}$ (see Fig. \ref{fig:optics}(c) for a diagram of the defined planes and Table \ref{tab:vars} for a list of constants in the model). We can solve for $d_{L3,F''}$ a function of $\Delta z$ by applying the imaging condition to $\mathbf{H}$,  meaning a ray input from the optical axis is output at the optical axis. 

\begin{equation}
\mathbf{H} 
\begin{bmatrix}
    0 \\
    \theta_{\mathrm{in}}
\end{bmatrix} 
=
\begin{bmatrix}
    1 & d_{L3,F''} \\
    0 & 1
\end{bmatrix}
\begin{bmatrix}
    A & B \\
    C & D
\end{bmatrix}
\begin{bmatrix}
    1 & f_{\mathrm{eye}}-\Delta z \\
    0 & 1
\end{bmatrix}
\begin{bmatrix}
    0 \\
    \theta_{\mathrm{in}}
\end{bmatrix}
=
\begin{bmatrix}
    0 \\
    \theta_{\mathrm{out}}
\end{bmatrix}
\label{eq:raymatrix}
\end{equation}

Solving \eqref{eq:raymatrix} for $d_{L3,F''}$:

\begin{equation}
    d_{L3,F''}=-\frac{Af_{\mathrm{eye}}-A\Delta z+ B}{C f_{\mathrm{eye}}-C\Delta z+D}
\label{eq:fundusimage}
\end{equation}

Finally, the lateral magnification of the system from $F$ to $F''$, $M_L$, is the first element of the system matrix, $H_{11}$.

\begin{equation}
    M_L = H_{11} = A + Cd_{L3, F''}
\label{eq:latmag}
\end{equation}

The size of a pixel in the object space, $R$, can be derived from the size of a sensor pixel, $\delta$, mapped to the object plane \cite{boominathan_phlatcam_2020}.
\begin{equation}
    R = \frac{d_{F'',D}}{d_{D,S}} * \frac{1}{M_L} * \delta
\label{eq:resolution}
\end{equation}

For our system, $A=1.340, B=-40.98, C=0.0125, D=0.3654$, and $\delta = 6.4$ microns. The distance from $F''$ to the diffuser is $d_{F'',D} = d_{L3,D}-d_{L3,F''}$, where $d_{L3,D}$ is the distance between L3 and the diffuser, which is equal to 111.6 mm in our system. Using these constants and equations we can compute the pixel size as a function of refractive error. To compute the maximum theoretical resolution, we determine the maximum frequency component passed by our PSF for each refractive error, $\epsilon$, above the noise floor of our camera from a computed modulation transfer function (MTF) (Fig. \ref{fig:vergence}(d)). 

The theoretical maximum resolution in an emmetropic eye with 60 diopters of power is computed to be approximately 12.5 lp/mm. This varies with defocus error due to a difference in magnification as described in \eqref{eq:fundusimage}-\eqref{eq:resolution} and ranges from 10 lp/mm in an eye with -12D myopia to 15 lp/mm in an eye with +12D hyperopia. This is within the Abbe diffraction limit of approximately 100 lp/mm imaging the retina through the 1.4 mm diameter aperture stop.

The FOV of an image can be estimated with the pixel size on the fundus corresponding to the defocus error and focal length of the eye. The FOV of fundus cameras is typically reported in terms of visual angle, which is the full cone angle of light collected from the pupil of the eye. From the paraxial Zemax model, we estimate the FOV of this system to be 40 degrees. 

\subsection{\textit{In vivo} imaging}
Alignment is crucial for capturing high-quality \textit{in vivo} fundus images. The illumination ring must align in the transverse plane to be concentric with the pupil, and the axial distance from the pupil to L1 must place the pupil conjugate to the diffuser. Alignment is performed by observing 1 frame per second video  with NIR illumination (740 nm, FWHM=30 nm) on undilated eyes in a dark room to limit pupil constriction, and reconstructions are shown live on a computer monitor. Once alignment is confirmed, three consecutive images with 40 ms exposures are taken using NIR (740 nm, FWHM=30 nm), green (550 nm, FWHM=60 nm), then blue (470 nm, FWHM=30 nm) illumination. This is within the approximately 200 ms pupil light reflex time \cite{bergamin_latency_2003}. RGB illumination was selected to enable color imaging to produce color similar to traditional fundus cameras. The specific wavelengths were chosen to emulate single-shot color imaging with a Bayer-filtered sensor. The \textit{in vivo} data was collected following a protocol approved by the Johns Hopkins University Institutional Review Board (IRB 00333664). Healthy volunteers signed an informed consent for to enroll in the study. The light safety of the device was evaluated with ISO 15004-2:2007. More detailed safety information is in Table \ref{tab:safety} in the appendix.

\section{Results}

\subsection{Resolution target}

\begin{figure}
    \begin{center}
    \begin{tabular}{c}
    \includegraphics[height=8cm]{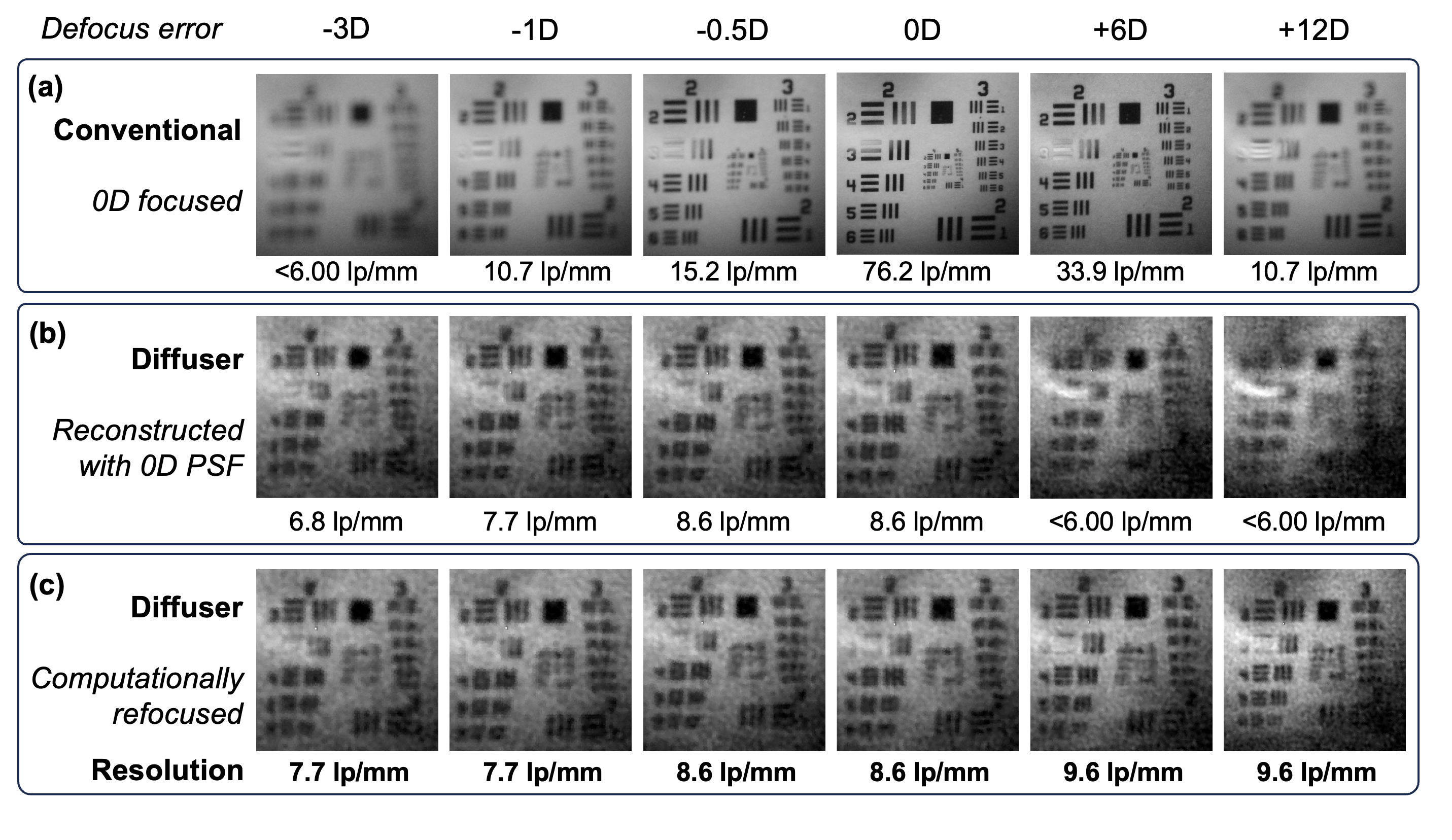}
    \end{tabular}
    \end{center}
    \caption{ \label{fig:restarget}
    Resolution evaluation of the system with a 1951 USAF resolution target. Defocus error was applied by adjusting the axial distance of the target behind a model cornea. The human eye-equivalent resolutions are shown, accounting for the difference in optical power between the human and model eye. (a) Cropped conventional camera images of the resolution target. The system was adjusted such that the target was in focus when axially positioned at 0D error. The resolution degrades sharply when the device is not focused. (b) Cropped diffuser images of the resolution target deconvolved with the 0D PSF. The resolution degrades with defocus error. (c) The resolution of the diffuser images is improved compared to (b) when the correctly corresponding PSF is used to deconvolve the same measurement.} 
\end{figure}

To characterize the resolution of the system across a range of defocus errors, images of a USAF 1951 resolution target (R1DS1P, Thorlabs) were taken with the computational fundus camera and the conventional camera (Fig. \ref{fig:restarget}). The resolution target was placed behind the model cornea at the focal distance. The captured image was deconvolved with Eq. \eqref{eq: inversefilter} using regularization strength $\mu=0.001$. The color channels were added together to eliminate differences in white balance between the separate images. To test the resolution on eyes with varying defocus errors, the resolution target was axially translated by hand and the image was computationally refocused. The translations, $\Delta z$, were kept under 6 mm, and 6 resolution test images were captured ranging from -3D to +12D of defocus error.

The resolution was estimated by examining the smallest set of line pairs that can be resolved. The mismatch between the power of the model eye cornea ($f_{\mathrm{model}}$ = 25.4 mm) and the human eye ($f_\mathrm{eye} \approx$  16.7 mm) is accounted for with a magnification correction. After determining the resolution in lp/mm on the resolution target behind the model cornea, the result is multiplied by a factor of $f_{\mathrm{model}}/f_\mathrm{eye}=1.52$ to estimate the expected resolution in a human eye. For comparison, conventional fundus photographs were taken with L2 fixed such that an object at 0D of error is in focus. 

The resolution of the computational (diffuser-based) imaging system is 7.7-9.6 lp/mm depending on the defocus error, as shown in Fig. \ref{fig:restarget}(c). The resolution of the system increases with more hyperopic defocus due to the increased magnification of the fundus, as expected (Fig. \ref{fig:vergence}(d)). The resolution is still poorer than the conventional lens-based image, which has a peak of 76.2 lp/mm for an emmetropic (0D) eye when the system has the focus set to 0D. However, this resolution quickly falls off for the conventional lens-based camera with defocus error applied. Meanwhile, the resolution of the refocused computational image remains in the 7.7-9.5 lp/mm range for all tested defocus errors (from -3D to +12D) despite not having to physically alter the system.

\subsection{Model eye}
Next, we tested our system on a commercial model eye (Ophthalmoscope Trainer, HEINE), which has an adjustable length to simulate defocus errors from -5D to +5D of defocus error. Diffuser images of the model eye were taken with the model eye adjusted to be -5D myopic, 0D emmetropic, and +5D hyperopic. For ground truth comparison, conventional fundus photographs were taken with the camera lenses adjusted and fixed such that an object at 0D of error is in focus. 
\begin{figure}
    \begin{center}
    \begin{tabular}{c}
    \includegraphics[height=16cm]{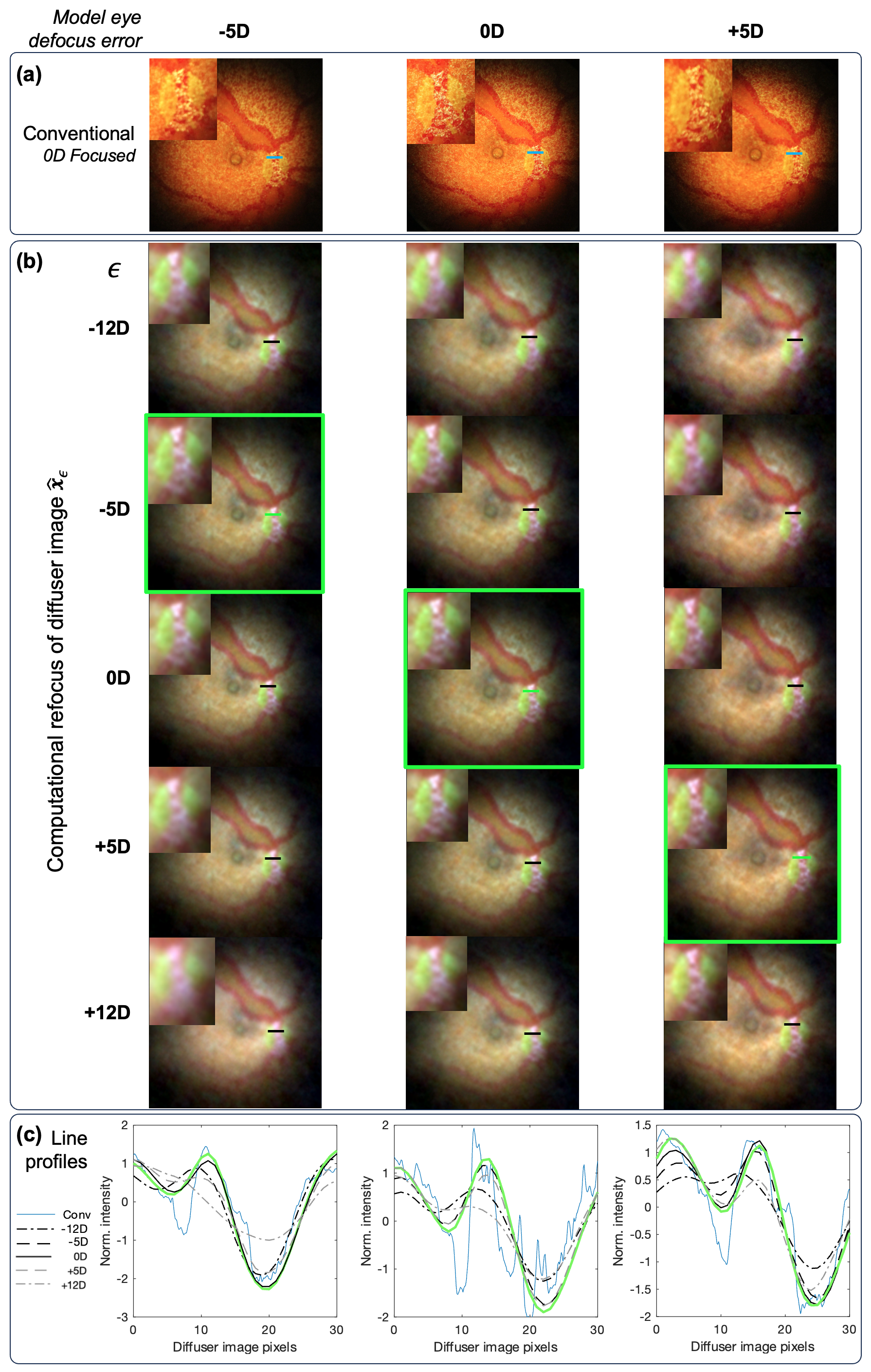}
    \end{tabular}
    \end{center}
    \caption{ \label{fig:modeleye}
    Computational refocusing in the HEINE model eye with defocus errors of -5D (first column), 0D (second column), and +5D (third column). (a) Conventional camera images with focus set to 0D. (b) Computationally refocused diffuser images using different PSFs. The diagonal indicated by green boxes contains the sharpest images when deconvolving with the corresponding PSF. The reconstructions are blurrier when the difference between the defocus error of the eye and the $\epsilon$ used for $h_\epsilon$ is greater. (c) Normalized profiles through the optic disk for each row. Among diffuser images, the contrast is the greatest in the correctly refocused image (green line for each row).} 
\end{figure} 

The results in Fig. \ref{fig:modeleye} show the reconstructions of the model eye. The reconstructions were computed with Eq. \eqref{eq: inversefilter} using $\mu=0.002$ for each color channel. The conventional image and diffuser image have a similar FOV, which is approximately the $40^{\circ}$ FOV inherent to the optical design of the Canon CR-DGi. The FOV was estimated using the theoretical pixel size computed with Eq. \eqref{eq:resolution}. Refocusing to the proper defocus error results in the sharpest reconstruction. The computational images in Fig. \ref{fig:modeleye}(b) can be computationally refocused post-capture, while a conventional image requires precise adjustment prior to image capture. Further work could improve the resolution of computational images with better PSF optimization.

\subsection{\textit{In vivo} images}
\begin{figure}
    \begin{center}
    \begin{tabular}{c}
    \includegraphics[height=10cm]{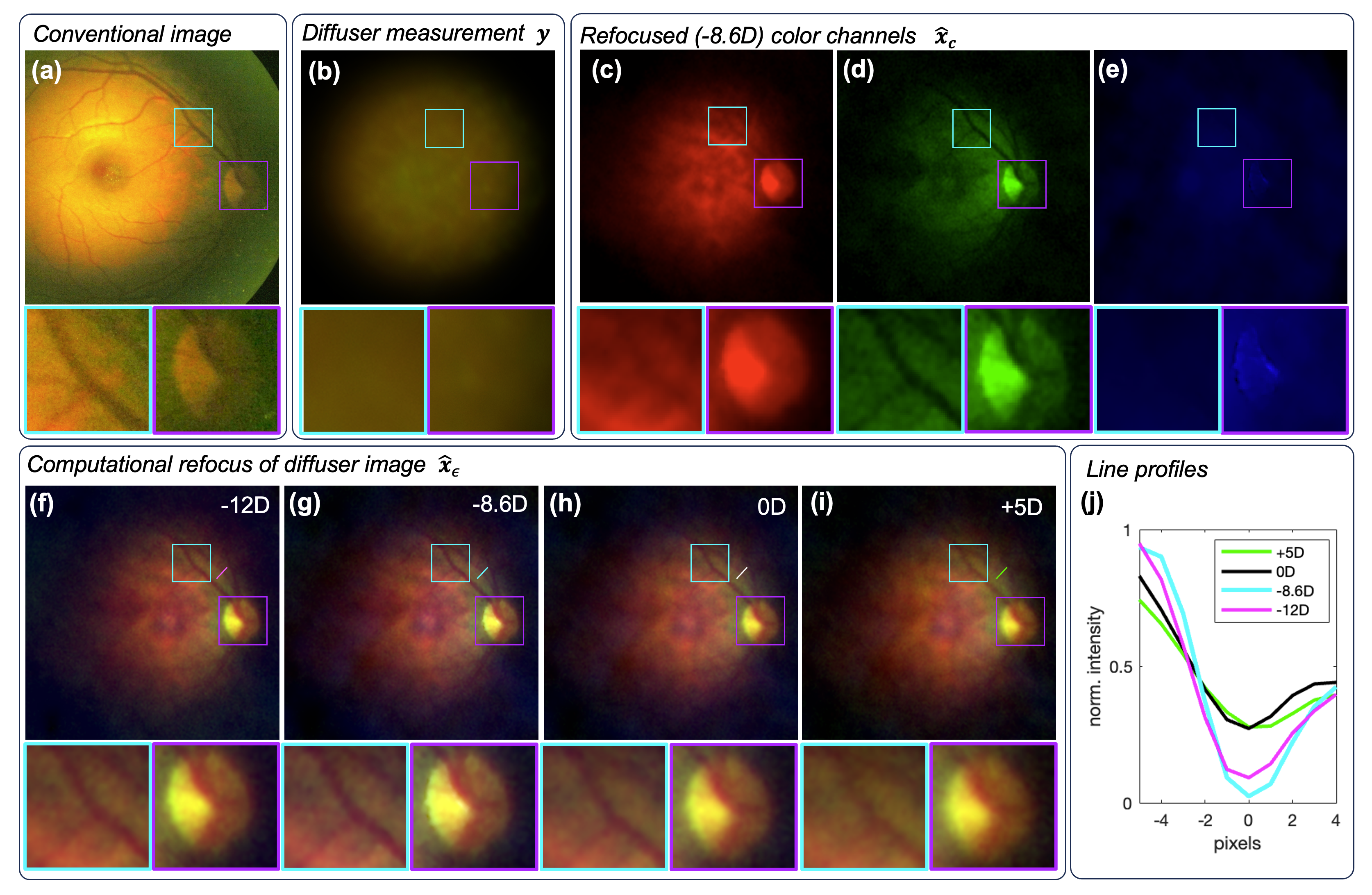}
    \end{tabular}
    \end{center}
    \caption{ \label{fig:invivofocus}
    (a) Conventional camera image of an \textit{in vivo} fundus. (b) Raw diffuser camera measurement of the fundus with RGB images combined. (c-e) Reconstructions of the individual red, green, and blue illumination measurements computationally refocused to -8.6 diopters. The green channel (d) provides the most vessel and optic disk contrast. (f-i) Computational refocusing of the combined RGB fundus image. The optic disk, blood vessels, and choroidal structure are observed. (g) The sharpest image corresponds to -8.6 diopters. (j) Line profiles of the green channel across the same vessel in (f-i). The -8.6D reconstruction has the greatest contrast and sharpest vessel boundaries. Cyan and purple box insets for each large field of view image are magnified and displayed below for comparison of vessel and optical disk contrast.} 
\end{figure}

We present the first diffuser-based computational images of an \textit{in vivo} fundus (Fig. \ref{fig:invivofocus}). Computational refocusing enables images to be resolved without prior knowledge of the patient's refractive error. The macula, optic disk, choroidal structure, and vessels are visible in the refocused reconstruction. The red, green, and blue color channels are created from three consecutively captured images with 40 ms exposures each and overlaid. The regularization strength was tuned separately for each color channel and was chosen to be $\mu=0.007$ for red, $\mu=0.002$ for green, and $\mu=0.02$ for blue. The green channel contains the most contrast due to the absorption of green light by blood vessels, thus requiring the least regularization. 

\section{Discussion}
\subsection{Resolution}
The resolution of the diffuser-based system was shown to be $\sim$7-10 lp/mm across a range of defocus errors (Fig. \ref{fig:restarget}). This is less than the maximum theoretical resolution of 10-15 lp/mm (Fig. \ref{fig:vergence}(d)). This is partly attributable to the scene being dense, making the inverse problem poorly conditioned \cite{adams_vivo_2022}, even with a low-noise, high bit-depth sensor. The resolution of the diffuser-based computational fundus camera is poorer than a perfectly focused conventional camera. However, the ability to computationally refocus the system to compensate for refractive error with no mechanical adjustments is a significant advantage that could simplify image capture. Unlike traditional fundus imagers, which struggle on eyes with significant refractive error and require mechanical focus adjustments, this approach enables post-capture correction. This may be particularly useful for applications where rapid and simple fundus imaging is desired. To improve the imaging resolution of the system while retaining the digital refocusing ability, in the future, the diffuser could be replaced with an element with a sparser PSF, such as a random microlens array. \cite{liu_fourier_2020}

\subsection{\textit{In vivo} imaging}
The results demonstrate that the diffuser-based computational fundus camera allows single-shot imaging of the fundus with no focusing required before image capture and can reconstruct images of the fundus over a large range of defocus errors. The \textit{in vivo} FOV is approximately 40 degrees, which was estimated using the theoretical pixel size for a 60 diopter eye with -8.6 diopters of myopia. This FOV is common for non-mydriatic fundus cameras used for diagnostic screening. Important clinical features, including the optic disk, macula, and larger blood vessels are resolved in both the model eye and \textit{in vivo} human eye. Computational refocusing increased the visibility and contrast of the blood vessels (Fig. \ref{fig:invivofocus}). 

There are some limitations to be addressed in future work. We did not have a fixation target, which made alignment and control of instrument-induced myopia difficult. Additionally, the resolution is worse than a clinical fundus camera (which typically have resolutions greater than 50 lp/mm). Future work will require increasing the resolution to ensure the diagnostic capabilities of the system. Regardless, this first demonstration of successful \textit{in vivo} diffuser-based imaging of the fundus is a significant step forward toward a future ocular imaging capabilities enabled by lensless imaging.

\subsection{Deconvolution methods}
\label{sec: disc: deconv}
Our deconvolution method is the same as Li et al. \cite{li_diffuser-based_2020} and is computationally simple. The Tikhonov inverse problem has a closed-form solution with no iteration required, making the reconstruction of a single image take approximately 200 ms and a focal stack of 500 images take less than 2 minutes. However, there is room for improvement with the use of more sophisticated regularizers. Other options are deep-learning models that may better reconstruct images by taking into account small amounts of spatial variance of the PSF\cite{yanny_deep_2022}. Numerous other deep-learning models have been developed for image reconstruction as well, such as GedankenNet by Huang et al. \cite{huang2023self}, which could be used in future work for computational ophthalmic imaging.

While deep learning reconstruction methods could likely improve some image reconstructions, there are several important drawbacks compared to the physical model-based inversion approach presented here. Deep learning methods generally require significant training data and large computing resources for training. More importantly, they risk hallucinations that could obscure or fabricate clinically meaningful features. The near-instantaneous reconstruction of a single image is particularly useful for \textit{in vivo} imaging because live feedback for the operator is necessary to verify alignment. Finally, we do not believe that the image reconstruction process is the greatest resolution-limiting factor in the system; rather, it is ultimately limited by the properties of the PSF and the noise floor of the camera, which provides a physical limit on spatial frequencies that can be recorded by the image sensor.

\subsection{Future work}

There are other approaches to computational fundus imaging using alternative optical encoding elements to enable digital refocusing. One approach is using coded aperture masks\cite{asif2016flatcam}. However, diffusers and other phase-mask elements are beneficial because they have greater light throughput, which is important in a light-constrained setting such as ophthalmic imaging. Another challenge to fundus imaging with multiplexing systems is that the fundus is a dense object which makes deconvolution of an extended PSF difficult \cite{adams_vivo_2022}. Future work using phase mask elements with sparser PSFs, such as random microlens arrays, could improve the resolution of the system \cite{liu_fourier_2020}. Other groups have found that random microlens arrays result in better resolution than holographic diffusers for computational imaging \cite{kuo_-chip_2020}, while the randomness alleviates the aliasing concerns associated with periodic microlens arrays. A specially designed phase mask for the computational fundus camera should balance the requirements of resolution and computational refocusing over the range of clinically relevant refractive errors. Specialized phase masks could also be designed to efficiently multiplex aberrometry information, including spherical and cylindrical errors, into the measurement. Other groups have developed methods for manufacturing custom phase masks using various lithography methods \cite{lee_design_2023,lee_fabrication_2022,zheng_close_2023}. McKay et al. have already explored diffuser-based aberrometry, which uses a similar configuration to this system \cite{mckay_large_2019}. 

Commercial low-cost handheld autorefractors have been developed that promise to expand access to eyeglass prescriptions \cite{durr_design_2015, durr_quality_2019}. We believe that phase mask-based fundus imaging could be leveraged to create a similar device that simplifies image capture and combines fundus imaging and autorefraction into a single measurement with a simple optical system. This would have great potential to reduce costs, increase throughput, and deskill the comprehensive eye examination. This could allow for single-shot aberrometry and fundus photography with no prior focusing required. Future reconstruction methods could involve synergistically reconstructing while performing wavefront sensing to improve the quality of the image and provide eyeglass prescriptions.

\section{Conclusion}
Access to eye diagnostic care could be improved with easier-to-use, higher-throughput diagnostic tools. This work builds upon the work of Antipa et al. \cite{antipa_diffusercam_2018} and Li et al. \cite{li_diffuser-based_2020} to create a diffuser-based fundus camera that enables post-capture computational refocusing of the fundus, circumventing one of the challenging steps in obtaining a high-quality fundus image. We demonstrate computational refocusing of \textit{in vivo} fundus images across a wide range of clinically relevant defocus errors. The resolution of the system is relatively consistent across many diopters of defocus error without the need for moving parts. The FOV is comparable to that of a standard non-mydriatic fundus camera. However, the resolution is worse than a conventional fundus camera, limited by the smoothness and spread of the diffuser PSF. This limitation could be addressed with better encoding elements, such as random microlens arrays. Lastly, there is an opportunity for future work that synergistically estimates aberrometry information and reconstructs a fundus image from a single measurement.

\pagebreak

\appendix    
\section{System parameters}
\begin{figure}[ht]
    \begin{center}
    \begin{tabular}{c}
    \includegraphics[height=8.5cm]{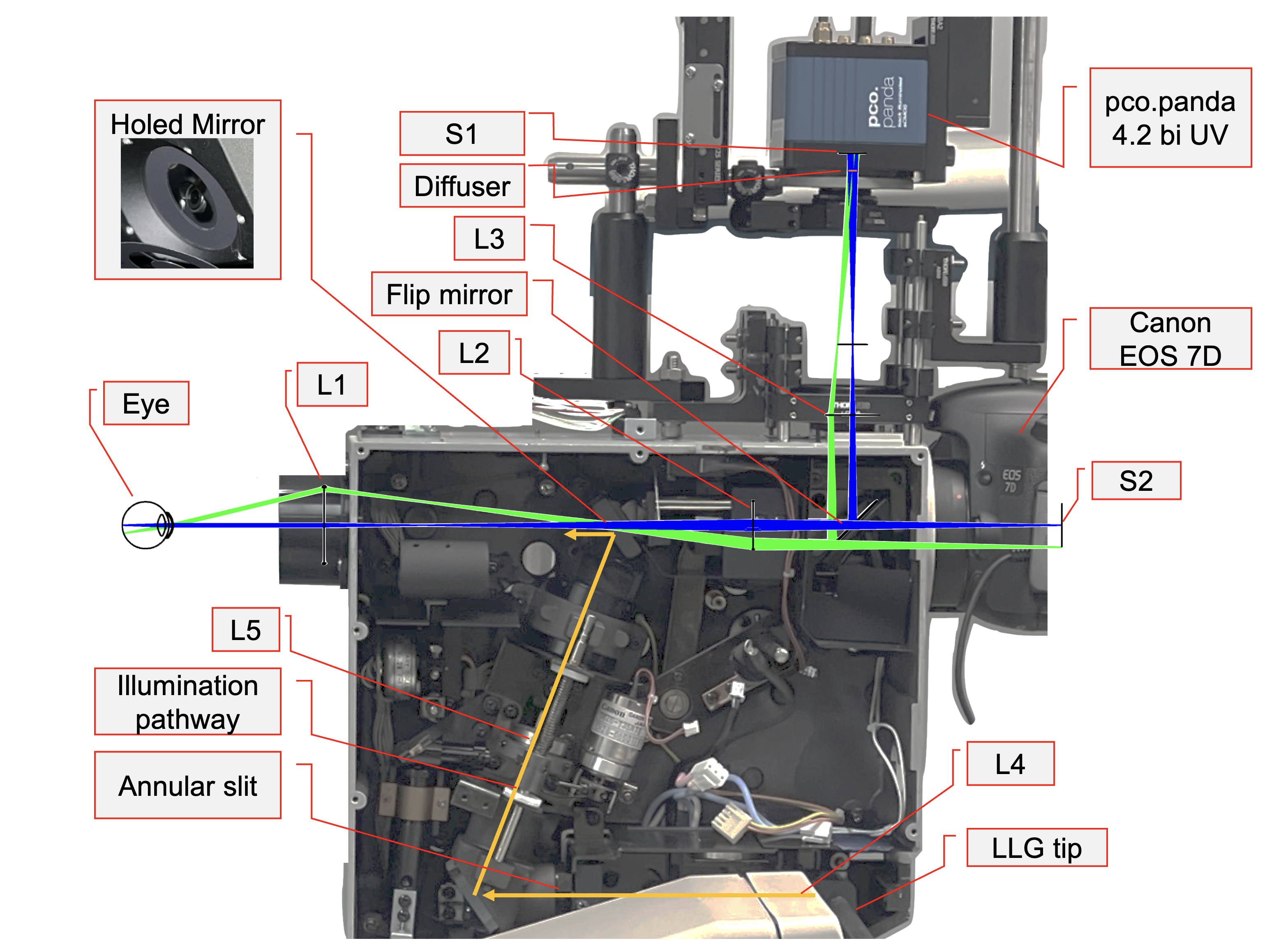}
    \end{tabular}
    \end{center}
    \caption{ \label{fig:systemphoto}
    Photograph of the system with labeled components. The diffuser camera is connected to the top of the commercial fundus camera, replacing the conventional IR alignment camera. The internal electrical components of the fundus camera were removed, and the inside of the Canon CR-DGi fundus camera was used for its optics. Zemax models of the diffuser imaging and conventional imaging path are overlaid. The entire system is mounted to a translating base that eases the alignment process. An ophthalmic chinrest is used to stabilize the patient's head (not pictured).} 
\end{figure}

Placement of the diffuser conjugate to the pupil is necessary for the PSF of the system to be shift-invariant. This is verified by translating the pinhole transversely behind the model cornea while observing the PSF created on the image sensor. The axial location of the diffuser was then fine-tuned with a translation stage until the PSF was maximally shift-invariant. During our testing, we also found that the diffuser-sensor distance ($d_{D,S}$) was important to optimize for the sharpest caustic pattern to form. This reduces the spread of the PSF which improves the conditioning of the inverse problem. We achieved this by placing the pinhole behind the model cornea at an emmetropic distance and translating the camera (pco.panda in Fig. \ref{fig:systemphoto}) axially while observing the sharpness of the caustic pattern formed on the image sensor.

The Zemax model shown in Fig. \ref{fig:optics} was made using thin lenses. The lenses that are part of the Canon CR-DGi (L1, L2, L4, and L5) do not have publicly accessible parameters, so they were estimated using measured distances between the conjugate planes of the system.  The components we used to build the system are in Table \ref{tab:parts}. The parameters of the Zemax model used to evaluate the system are in Table \ref{tab:vars}.

\begin{table}[ht!]
    \caption{Optical components in the computational imaging pathway.}
    \label{tab:parts}
    \begin{center}
    \begin{tabular}{|l|l|l|l|l|l|} 
    \hline
    \rule[-1ex]{0pt}{3.5ex} Part & Description & Part no. & Manufacturer & Focal len. (mm) & Diameter (mm) \\
    \hline\hline
    \rule[-1ex]{0pt}{3.5ex}  L1 & Objective &   & Canon & 42$^*$ & 50.8 \\ 
    \hline
    \rule[-1ex]{0pt}{3.5ex}  L2 & Relay     &   & Canon & 75$^*$ & 25.4 \\ 
    \hline
    \rule[-1ex]{0pt}{3.5ex}  L3 & Relay 2 & AC254-100-A & Thorlabs & 100 & 25.4 \\ 
    \hline
    \rule[-1ex]{0pt}{3.5ex}  D & Diffuser & $\#$47-988 & Edmund Optics & $\approx$11$^*$ & 25.4 \\ 
    \hline
    \rule[-1ex]{0pt}{3.5ex}  C & Model cornea & LB1761-A & Thorlabs & 25.4 & 25.4 \\ 
    \hline
    \rule[-1ex]{0pt}{3.5ex}  C & Model eye&  C-000.33.010 & HEINE & 18 & 8 \\
    \hline\hline
    \multicolumn{3}{l}{\footnotesize * represents a computed or estimated value}
    \end{tabular}
    \end{center}
\end{table}

\begin{table}[ht!]
    \caption{List of variables in system model.}
    \label{tab:vars}
    \begin{center}
    \begin{tabular}{|l|l|l|} 
    \hline
    \rule[-1ex]{0pt}{3.5ex} Variable & Description & Value \\
    \hline\hline
    \rule[-1ex]{0pt}{3.5ex}   $f_{\mathrm{eye}}\,^*$ & \textit{In vivo} eye & 100/6 mm \\ 
    \hline
    \rule[-1ex]{0pt}{3.5ex}  $f_{\mathrm{HEINE}}$ & HEINE model eye & 18 mm \\ 
    \hline
    \rule[-1ex]{0pt}{3.5ex}  $f_{\mathrm{model}}$ & Model cornea & 25.4 mm \\ 
    \hline
    \rule[-1ex]{0pt}{3.5ex}  $d_{P,L1}$ & Pupil to L1 dist. & 70.85 mm \\ 
    \hline
    \rule[-1ex]{0pt}{3.5ex}   $f_{L1}\,^*$ & Objective lens & 46.2 mm \\ 
    \hline
    \rule[-1ex]{0pt}{3.5ex}  $d_{L1,HM}$ & L1 to HM dist. & 132.8 mm \\
    \hline
    \rule[-1ex]{0pt}{3.5ex}  $d_{HM,L2}$ & HM to L2 dist. & 68.75 mm \\
    \hline
    \rule[-1ex]{0pt}{3.5ex}   $f_{L2}\, ^*$ & Relay lens & 75 mm \\ 
    \hline
    \rule[-1ex]{0pt}{3.5ex}   $d_{L2,L3}$ & L2 to L3 dist. & 96.75 mm \\ 
    \hline
    \rule[-1ex]{0pt}{3.5ex}   $f_{L3}$ & Diffuser relay lens & 100 mm \\ 
    \hline
    \rule[-1ex]{0pt}{3.5ex}   $d_{L3,D}$ & L3 to D dist. & 111.6 mm \\ 
    \hline
    \rule[-1ex]{0pt}{3.5ex}   $d_{D,S}$ & Diffuser sensor dist. & 11.0 mm \\ 
    \hline\hline
    \multicolumn{3}{l}{\footnotesize * represents a computed or estimated value}
    \end{tabular}
    \end{center}
\end{table}

\section{Safety}
The device was evaluated for light safety according to ISO 15004-2:2007, which specifies light hazard protection for ophthalmic devices. This protocol specifies the maximum irradiance on the cornea and fundus, considering the thermal hazard, aphakic photochemical hazard, and ultraviolet hazard. For each wavelength band used, the spectrum was recorded with a spectrometer and the optical power tests weighted for all hazards were measured as prescribed by the ISO protocol. For each wavelength, including the 740 nm NIR/red used for alignment, the device was determined to be Group 1, indicating that it is non-hazardous for human use. The maximum irradiances for the different wavelengths are shown in the Table \ref{tab:safety}. In practice, the power used for each color band was restricted to less than these values.

\begin{table}[ht!]
    \caption{Summary of the maximum irradiances for the wavelengths used by the computational fundus camera.}
    \label{tab:safety}
    \begin{center}
    \begin{tabular}{|p{0.2\linewidth}|p{0.15\linewidth}|p{0.15\linewidth}|p{0.15\linewidth}|p{0.15\linewidth}|} 
    \hline
    \rule[-1ex]{0pt}{3.5ex}  Wavelength (nm) & Total power \newline output \newline (mW) & Anterior segment irradiance (mW/cm\(^2\)) & Fundus \newline irradiance (mW/cm\(^2\)) & Passes \newline all ISO limits for Group 1 \\ 
    \hline\hline
    \rule[-1ex]{0pt}{3.5ex}  470 (blue) & 5.90 & 74 & 7.4 & Yes \\
    \hline
    \rule[-1ex]{0pt}{3.5ex}  550 (green) & 5.66 & 71 & 7.1 & Yes \\ 
    \hline
    \rule[-1ex]{0pt}{3.5ex}  740 (NIR/red) & 3.60 & 45 & 4.5 & Yes \\ 
    \hline
    \end{tabular}
    \end{center}
\end{table}
\pagebreak
\subsection*{Disclosures}
NJD is listed as co-inventor on a provisional patent application assigned to Johns Hopkins University that is related to the technologies described in this article. They may be entitled to future royalties from this intellectual property.

\subsection* {Code, Data, and Materials Availability} 
Code and data used in this manuscript are available in the Johns Hopkins Research Data Repository \cite{T17LHBOF_2025}. Code is also available in a GitHub repository at \url{https://github.com/DurrLab/Computational-Funduscopy}.

\subsection* {Acknowledgments}
This work is supported by the Wilmer Eye Institute Pooled Professor Fund and the unrestricted grant from Research to Prevent Blindness (RPB).

\pagebreak
\bibliography{article}   
\bibliographystyle{spiejour}   


\vspace{2ex}\noindent\textbf{Corey Simmerer} is a PhD candidate in the department of biomedical engineering at the Johns Hopkins University School of Medicine. He received his BSE degree from Duke University in biomedical engineering and electrical and computer engineering in 2021, and his MS degree from Johns Hopkins in biomedical engineering in 2024. His research interests include biophotonics and computational imaging.

\vspace{2ex}\noindent\textbf{Marisa Morakis} is a PhD candidate in the department of biomedical engineering at the Johns Hopkins University School of Medicine. She received her BS degree from Bucknell University in 2020. Her research interests include biophotonics and computational imaging.

\vspace{2ex}\noindent\textbf{Lei Tian} is an associate professor of electrical and computer engineering at Boston University. He received his MS and PhD from MIT in 2010 and 2013, respectively. His current research interests include computational imaging and sensing, imaging in scattering media, phase retrieval, and neurophotonics.

\vspace{2ex}\noindent\textbf{Lia Gomez-Perez} is a PhD student in medical physics in the Harvard-MIT Health Sciences and Technology program. She received her BS in biomedical engineering from The Ohio State University in 2023, and did undergraduate research at Johns Hopkins University in 2022 as a part of the Computational Sensing \& Medical Robotics REU program.

\vspace{2ex}\noindent\textbf{T.Y. Alvin Liu} is an assistant professor of ophthalmology at the Wilmer Eye Institute at Johns Hopkins University. He received his BA from Cornell University and MD from Columbia University, and did his fellowship training and residency at Johns Hopkins University. His current research interests include artificial intelligence and ophthalmic imaging.

\vspace{2ex}\noindent\textbf{Nicholas J. Durr} is an associate professor of biomedical engineering at Johns Hopkins University. He received his BS degree in electrical engineering and computer science from UC Berkeley in 2003 and MS and PhD degrees in biomedical engineering from UT Austin in 2007 and 2010, respectively. His current research interests include medical devices, machine learning, and biophotonics. He is a member of SPIE.

\listoffigures
\listoftables

\end{spacing}
\end{document}